# Three-dimensional graphene on a nano-porous 4H-SiC backbone: a novel material for food sensing applications

*Stefano Veronesi[1,*]  Ylea Vlamidis[1,2]  Letizia Ferbel[1]  Carmela Marinelli[2]  Chiara Sanmartin[3] Isabella Taglieri[3]  Georg Pfusterschmied[4]  Markus Leitgeb[4]  Ulrich Schmid[5]  Fabio Mencarelli[3] Stefan Heun[1]*

[1]NEST, Istituto Nanoscienze-CNR and Scuola Normale Superiore, Piazza S. Silvestro 12, 56127 Pisa, Italy Email Address: stefano.veronesi@nano.cnr.it

[2]Department of Physical Science, Earth, and Environment, University of Siena, Via Roma 56, 53100, Siena, Italy

[3]Department of Agriculture, Food and Environment Science, University of Pisa, Pisa, Italy

[4]Institute of Sensor and Actuator Systems, TU Wien, Vienna 1040, Austria

BACKGROUND: Sensors which are sensitive to volatile organic compounds and thus able to monitor the conservation state of food, are precious because they work non–destructively and allow to avoid direct contact with the food, ensuring hygienic conditions. In particular, the monitoring of rancidity would solve a widespread issue in food storage.

RESULTS: The sensor discussed here is produced utilizing a novel three-dimensional arrangement of graphene, which is grown on a crystalline silicon carbide (SiC) wafer previously porousified by chemical etching. This approach allows a very high surface-to-volume ratio. Furthermore, the structure of the sensor surface features a large amount of edges, dangling bounds, and active sites, which make the sensor, on a chemically robust skeleton, chemically active, particularly to hydrogenated molecules. The interaction of the sensor with such compounds is read out by measuring the sensor resistance in a four wire configuration. The sensor performance has been assessed on three hazelnut samples: sound hazelnuts, spoiled hazelnuts, and stink bug hazelnuts. A resistance variation of about $\Delta R$ = 0.13 ± 0.02 Ω between sound and damaged hazelnuts has been detected.

CONCLUSIONS: Our measurements confirm the ability of the sensor to discriminate between sound and damaged hazelnuts. The sensor signal is stable for days, providing the possibility to use this sensor for the monitoring of the storage state of fats and foods in general.

## 1 Introduction

Food traceability, quality control, and contamination issues represent hot topics to improve food production, distribution, and consumption. The possibility to follow the preservation state of food, in order to ensure the best quality of the products, minimize food losses, and take care of the consumers health is the driving force for a large body of research work. Indeed, a proper monitoring of the food conservation state has a strong impact on both the health of consumers and the food waste issue. This issue has been included within the 17 goals of the United Nations (UN) Sustainable Development, in particular in the goal 12 [1]. The UN estimates that more than 13% of the food is lost from farm to processing, and a further 17% at the consumer level. Moreover, this trend is unchanged from 2016 to 2021, far from the target to reducing losses by 50% by 2030. Besides, waste of food produces an additional contribution to the global warming, as well. Research is strongly involved in the effort to mitigate waste of food and to protect the health of consumers, developing new and improved sensors and studying intelligent packaging. Nevertheless, the use of food sensors remains limited. Importantly, the development of sensors able to monitor the degradation of fats present in food, both from natural occurrence or added during the food processing, is in an early stage despite the large number of possible applications. Volatile Organic Compounds (VOCs) are responsible of flavors and aromas of plants and fruits, and their oxidative processes are related to an alteration of taste and odor [2]. In addition, they are a by-product of the fat oxidation/degradation that can occur during the food lifetime, resulting in changes in the sensory perception of the product. Therefore, a way to monitor the degradation of fats in food is to detect VOCs. The efficiency of detection is crucial for a VOCs sensor. High detection efficiency can be obtained by maximizing the probability that a target analyte meets an active site of the detector and interacts with it. This goal can be achieved by increasing the efficiency of the active sites and/or by maximizing the number of active sites per unit of area and/or by increasing the useful sensor surface. Therefore, the availability of materials with a large surface-to-volume ratio represents a benefit. A largely utilized platform in developing detectors is graphene, which has been used to implement optoelectronic applications [3], wearable electronics [4, 5], sensors [5, 6, 7, 8, 9, 10], and biosensors [11]. The outstanding properties of graphene can be further tailored by chemical functionalization. However, in many fields such as catalysis [12], supercapacitors [13, 14], water filtration [15, 16], and drug delivery [17], a three-dimensional structure increases the surface above average compare to the 2D counterparts. Three-dimensional structures are also used to realize high performance electrodes [18, 19], gas detection sensors [20], and battery cathodes [21].

A three-dimensional arrangement of graphene combines the outstanding properties of graphene with the requirement of a large active surface area in developing high sensitivity detectors. The sensor presented here is based on a novel three-dimensional graphene arrangement (3DG in the following) that was recently developed [22]. 3DG samples are realized via growth of epitaxial graphene on the Si-face of a 4HSiC wafer that has been previously porousified via photoelectrochemical etching [23, 24, 25]. This material has already been utilized for hydrogen storage purposes and demonstrated catalytic properties, which allowed to chemisorb, for the first time in a pristine graphene material, hydrogen atoms starting from molecules [26].

Hazelnut (Corylus avellana L.) is a dried fruit largely considered throughout the world as relevant raw material for chocolate, confectionery and bakery industries [27]. Turkey, Italy, Spain, and the USA are the most important producers (FAO, 2006), even if new producers, located in the Southern Hemisphere, like Australia and Chile, are emerging [28, 29]. As 90% of the production is processed, the commercial quality of hazelnuts is mainly determined by the requirements of the confectionery industry [30] and, in this sense, good analytical methods for quality sorting and provenance identification are required. As with most dried fruits, the lipid matrix is very sensitive to chemical changes which affect the qualitative characteristics and for this reason techniques of storage have been proposed in recent years [31, 32]. The research of non–destructive sensors to discriminate hazelnuts has been performed for a long time [33]. Unfortunately, no reliable results to apply on line and on time have been reached. Here, a sensor from 3DG has been used to perform measurements on a blank sample and on three further samples, i.e., sound hazelnuts, spoiled hazelnuts, and stink bug hazelnuts, respectively. The sensor resistance $R_s$ is used as a sensitive and stable signal. A specific surface chemical functionalization to improve sensor performance and selectivity is discussed.

## 2 Sensor fabrication

The starting material used for sensor fabrication are Nitrogen–doped 4H-SiC wafers with a thickness of 350 $\mu$m and a bulk resistivity of 0.106 $\Omega$cm, oriented 4° off-axis with respect to the (0001) basal plane, corresponding to the Si-face. The wafers are porousified through a metal-assisted photochemical etching (MAPCE), followed by a photo-electrochemical etching (PECE), according to the procedure described in Ref. [24]. The main steps of the porousification process are schematically reported in **Figure 1**(a). The porousified wafers are cut into pieces with dimensions 2 mm × 7 mm, which have then been utilized for the epitaxial graphene growth.

The graphene growth is performed via SiC thermal decomposition in an ultra high vacuum (UHV) environment (see Figure 1(b)) at a base pressure $< 1 \times 10^{-10}$ mbar, annealing the sample at about 1370°C for 3 minutes. The growth procedure is reported in detail in Ref. [22]. After the growth, the graphene quality has been verified by Raman spectroscopy, reported in Figure 1(c). The 2D Raman peak has a FWHM of 54 cm$^{-1}$, slightly larger than expected for monolayer graphene. The ratio I(2D)/I(G) is 0.86, lower than the usual value for epitaxial graphene. As shown and discussed in detail in Ref. [22], both FWHM and I(2D)/I(G) values are mainly due to the strain-doping effect [34, 35, 36] of the graphene grown on the porous layer of the 4H-SiC wafer. The presence of D and D' peaks indicates the presence of defects, mainly related to the reduced graphene grain dimensions in these porous structures.

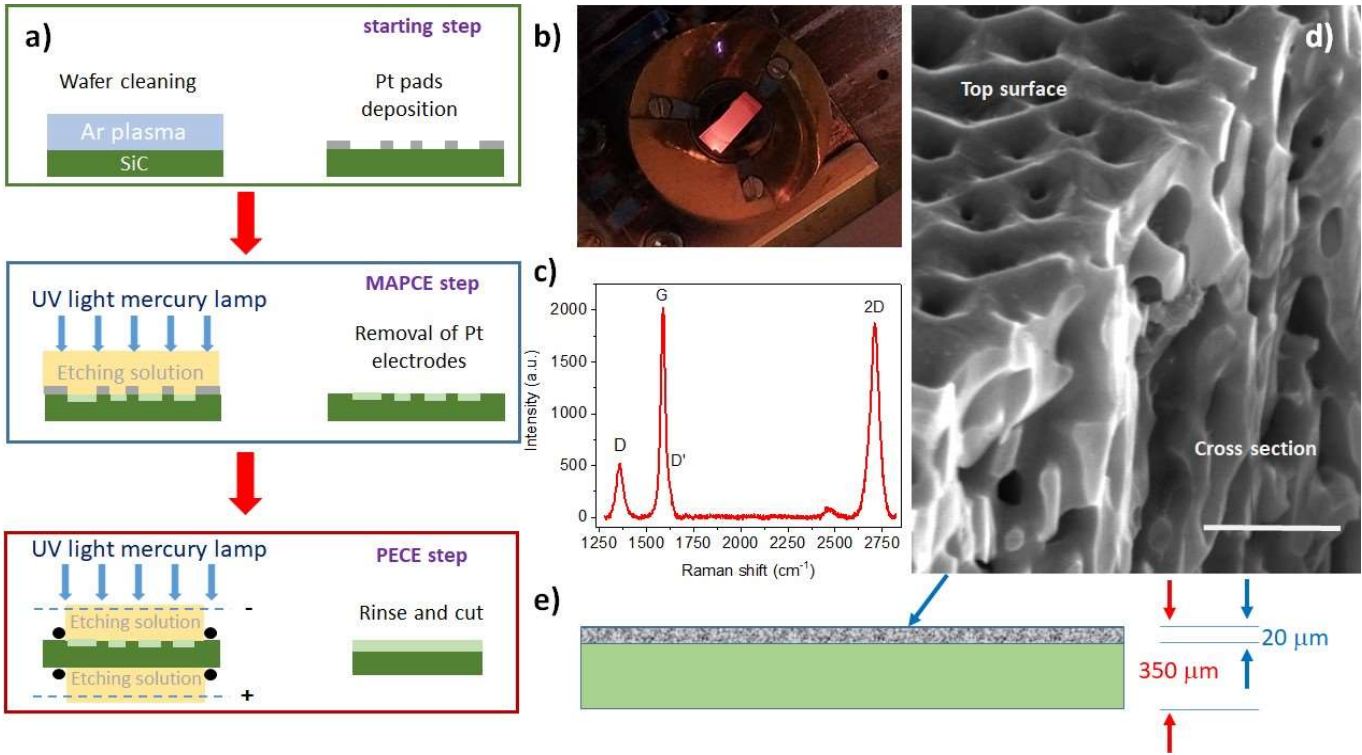

Figure 1: (a) Scheme af the MAPCE-PECE porousification process. (b) A porous 4H-SiC sample during the growth of epitaxial graphene. (c) Raman spectrum from the top surface of a 3D-graphene sensor. Shape and intensity ratio of the 2D and G peaks demonstrate the good quality of the graphene. (d) SEM image taken at 5 kV (beam current 11.7 pA) on the cross sectional edge of a 3DG sensor mechanically cleaved. The scale bar corresponds to 300 nm. (e) Sketch of a 3DG sensor.

In order to assess the homogeneity and the quality of the graphene inside the porous structure, a few sacrificial samples have been cleaved, and cross sections were investigated by scanning electron microscopy (SEM) (see Figure 1(d)) and Raman spectroscopy, as well, as reported in Ref. [26]. The etching parameters produce a porous layer of about 20 µm thickness, as sketched in Figure 1(e).

As a sensor signal, the variation of the resistance of the 3DG is measured in a four–wire (4W) configuration. An alternating current of $I$ = 1 µA is supplied to the sensor while the voltage drop $V_{4W}$ is measured with a lock-in amplifier. This technique allows a sensitive measurement of the sensor resistance $R_s$ with negligible impact from the contact resistance.

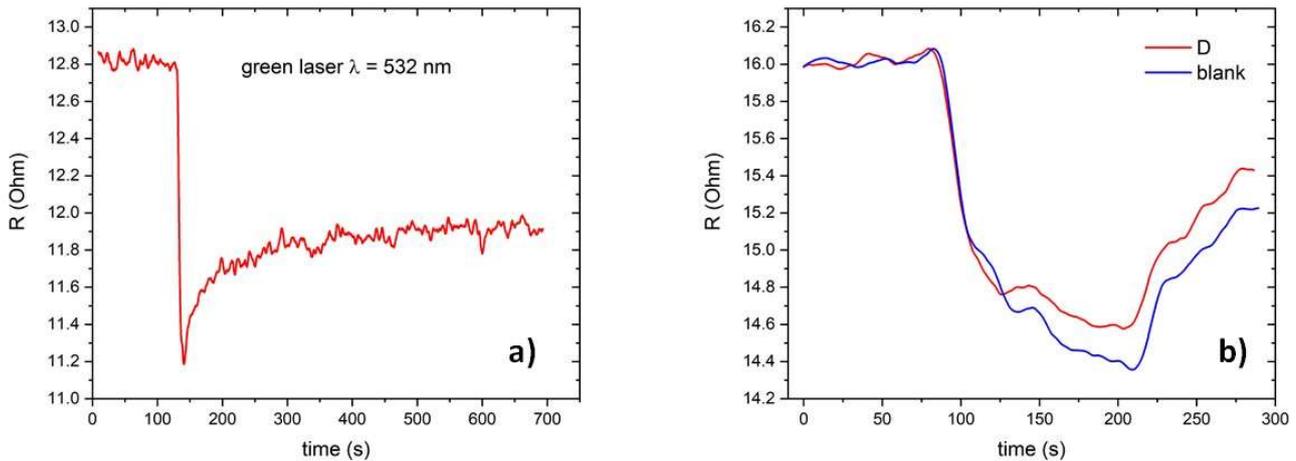

Figure 2: (a) Variation of the sensor resistance upon illumination with a green laser for 10 s. (b) Variation of the sensor resistance upon hydrogenation (red line). Blue line: blank experiment without hydrogen.

Before starting the evaluation of the sensor in detecting VOCs, its ability to respond to simpler physical stimuli has been verified. The first test of the sensor has been performed by illuminating it with a green laser. Upon illumination under UHV conditions, an increase in the current due to photoelectrons is detected, while the sensor resistance drops by 15%, as shown in **Figure 2**(a). We want to underline that the photon energy is lower than the SiC bandgap, therefore the absorption must be due to the graphene top layer. This demonstrates the ability of the sensor to detect a photocurrent. We add that light detection is not the focus of this work, and therefore we have not performed a calibration to quantitatively evaluate the sensor sensitivity as a function of wavelength. In a second measurement, the sensor was exposed to a flux of atomic hydrogen (exposure pressure $10^{-7}$ mbar) under UHV conditions. For details, see Ref. [26]. Hydrogen molecules are cracked with a Tectra hydrogen cracker via thermal dissociation of the hydrogen molecules on a hot tungsten tube at 1700 K. Therefore, the sensor is heated by the cracker, resulting in a signal even without hydrogen supply, as shown in Figure 2(b) (blank, blue line). This shows that the sensor is also an efficient thermometer. In the hydrogenation experiment, however, besides the thermal variation, a different signal dynamics is observed with respect to the blank, demonstrating the ability of the sensor to detect the hydrogen uptake, even at low exposure pressure.

## 3   3D-graphene as food storage state sensor

Since VOCs have a relatively low molecular weight and a high vapor pressure, they are ideal targets for gas phase detection. The ability of the sensor to detect VOCs related to the degradation of the storage state of hazelnuts has been demonstrated utilizing three hazelnut batches. The first was made by perfectly preserved hazelnuts, the second by spoiled hazelnuts, and the third by stink bug hazelnuts. The experiments were performed in an air-tight glass container with the sensor mounted on the bottle cap. During experiments, the glass container is closed, to avoid exchange of air from the inside to the outside, and vice-versa.
In the first series of experiments, the experimental protocol adopted to assess the sensor performance contemplated four measurements. The signal is acquired for a long time, some days, in order to understand the influence of the environment on the signal and to evaluate a possible saturation effect during the measurements. The first measurement is a blank experiment. The glass container is empty and closed, and thus the sensor is just exposed to air. Next, we performed three measurements with sound, spoiled, and stink bug hazelnuts, respectively. An amount of hazelnuts corresponding to about 60% of the volume is introduced in the glass container. This first series of experiments has been performed in an air-conditioned room, without any further active control of the sensor temperature. The results of the complete series of measurements performed on blank, sound, spoiled, and stink bug hazelnuts are shown in **Figure 3**. As can be seen, the data seems to show a small increase in the sensor resistance exposed to sound hazelnuts, but the error bars of the sensor resistance exposed to different environments are so large that it is hard to draw further conclusions.

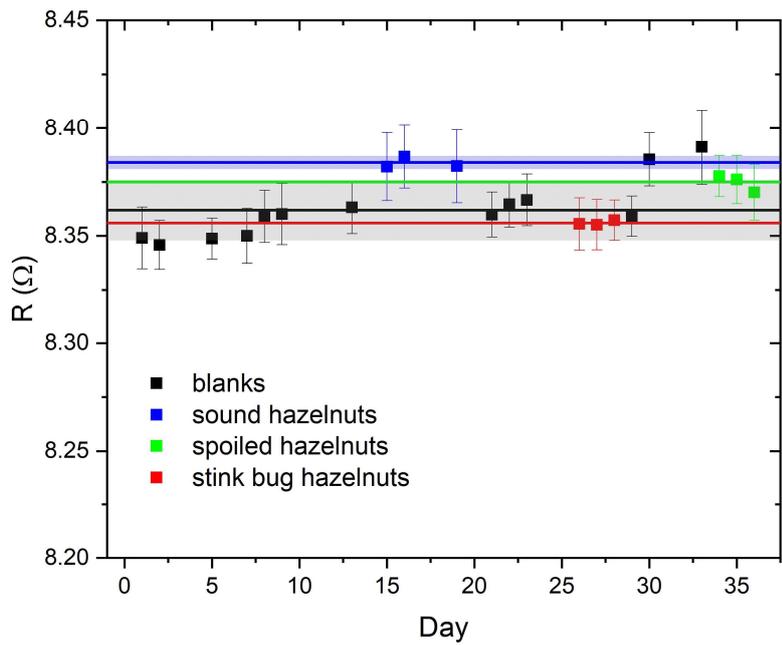

Figure 3: Sensor resistance in a series of measurements performed for more than a month. The black points represent blanks, blue points refer to the sensor exposed to sound hazelnuts, green points to spoiled hazelnuts, and red points to stink bug hazelnuts. Error bar of each data point is the standard deviation of the respective measurement. Color lines are the average of each population, and blueish and greyish areas the standard deviation for healthy hazelnuts and blank, respectively.

Analyzing the data set more closely, the origin of these fluctuations becomes clear. In spite of the fact that the experiment is performed in an air-conditioned room, the residual daily temperature fluctuations induce a signal variation that is much larger than the single measurement accuracy. The effect is highlighted in **Figure 4** where the circadian temperature fluctuation of the laboratory is evident. Each point in the figure is an average over 15 minutes of acquisition, and the whole data set spans for about 48 h. The standard deviation of a single average is about 0.006 Ω while the oscillation due to temperature fluctuations is around 0.045 Ω, nearly an order of magnitude larger. Figure 4 is a clear indication that the sensor temperature must be kept constant in order to avoid measurement fluctuations and to increase the overall accuracy.

The ability of the sensor to detect a temperature variation was already shown in Figure 2(b) when the sensor was heated by the hydrogen cracker. Here, in the measurements with the hazelnuts, the parasitic, temperature-induced signal is greater than the target signal. Thus, an active feedback is required to keep the sensor temperature constant. This is achieved with a resistive heater (50 Ω, 1 W), a K-type thermocouple which reads the sensor temperature, and a temperature controller (LakeShore Model 331). The transition from room temperature to the set point is shown in **Figure 5**. The sensor temperature was set to 40° C, and the long term stability is better than 0.07° C in 60 hours of acquisition. For comparison, the figure shows the readout variation induced by a temperature fluctuation of ±0.5° C (red bar). Furthermore, the heater allows to be operated up to 200° C. This feature can be utilized to periodically clean and degas the sensor, if required. We add that the sensor material itself is stable in temperature up to at least 900° C, allowing to employ the sensor in a wide range of environmental conditions. After this, a second series of measurements has been performed, in the same configuration of those shown in Figure 3, but with the addition of the sensor temperature stabilization. The data set obtained from this second experiment is shown in **Figure 6**. Working at a constant sensor temperature has clearly improved sensor performance. Now the sensor resistance exposed to sound hazelnut is clearly greater than the blank resistance and that of the harmed hazelnut samples. The sensor resistance increases or decreases depending on the interaction with the target molecules. If in the interaction an electron is released, then the resistance decreases, while if during the interaction an electron is bound, the resistance increases. Therefore, as the oxidative processes change the VOCs composition, the sensor resistance changes consequently. These results are summarized in Table 1. Indeed, both spoiled and stink bug hazelnuts produce similar VOCs changes [37] and hence result in very similar sensor resistance. Therefore, this sensor shows, for the first time to the best of our knowledge, the ability to discriminate values between sound and damaged hazelnuts.

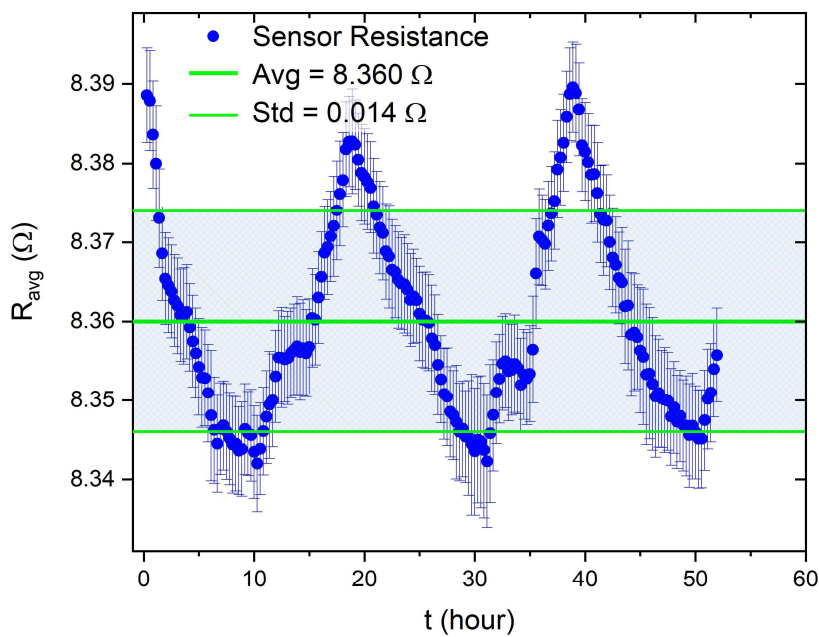

Figure 4: Sensor resistance variation during a two day long acquisition. The main oscillation is due to the residual circadian temperature oscillation of the air-conditioned laboratory. Thick green line is the overall average, and the thinner green lines the related standard deviation at the border of the blueish area.

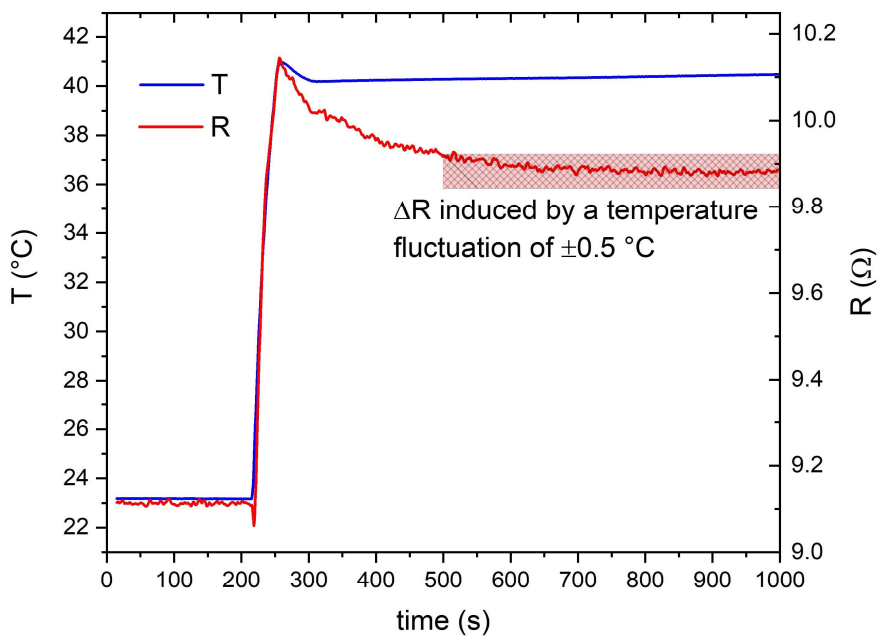

Figure 5: Variation in sensor temperature and resistance during 15 minutes, including the switch on of the temperature stabilization. The reddish area visualizes the effect of a ±0.5° C temperature fluctuation on the resistance readout.

Table 1: Sensor resistance measured for the blank and the healthy, spoiled, and stink bug hazelnuts. Each value is the average of the corresponding data, and the reported error is their standard deviation.

|  | blank | healthy hazelnuts | spoiled hazelnuts | stink bug hazelnuts |
| --- | --- | --- | --- | --- |
| Measured sensor resistance (Ω) | 9.86 ± 0.02 | 9.992 ± 0.016 | 9.83 ± 0.03 | 9.87 ± 0.04 |
| ΔR with respect to blank (Ω) | 0 | 0.132 ± 0.036 | -0.03 ± 0.05 | 0.01 ± 0.06 |

## 4  Discussion

The detection of VOCs, in particular, allows the control of many degradation processes responsible for the rancidity of fat-containing foods. Nuts naturally contains a lot of VOCs such as alcohols, aldehydes, ketones, esters, and ethers [38]. The concentration of aldehydes and alcohols are commonly related to the oxidation of fatty acids [38] and can be utilized to monitor the deterioration of nuts. In particular, the rancid flavor is related to a pool of molecules whose most relevant compounds are hexanal and nonanal. The sensor that we have developed is able to discriminate between samples of healthy hazelnuts and samples in which degradation occurred. The sensor is sensitive to temperature, and we have shown that it requires a careful temperature stabilization to work properly. With this temperature stabilization, however, data fluctuation is dramatically reduced, allowing to clearly discriminate between healthy and harmed fruits. Even if in the present work the active temperature control has been performed with standard research laboratory equipment, it is easy to integrate an on–chip temperature control in future devices.

Our results are promising and open the possibility to develop a sensitive device to monitor the storage state of hazelnuts, presumably working for different kinds of nuts. In the present investigation, we have not performed a specific functionalization to increase the sensitivity and selectivity of the sensor. Indeed, the pristine sensor is already able to discriminate between healthy and damaged hazelnuts. The difference between the sensor resistance exposed to healthy fruits and to damaged hazelnuts is more than 6 times the standard deviation of the measurement, giving a high confidence to their discrimination.

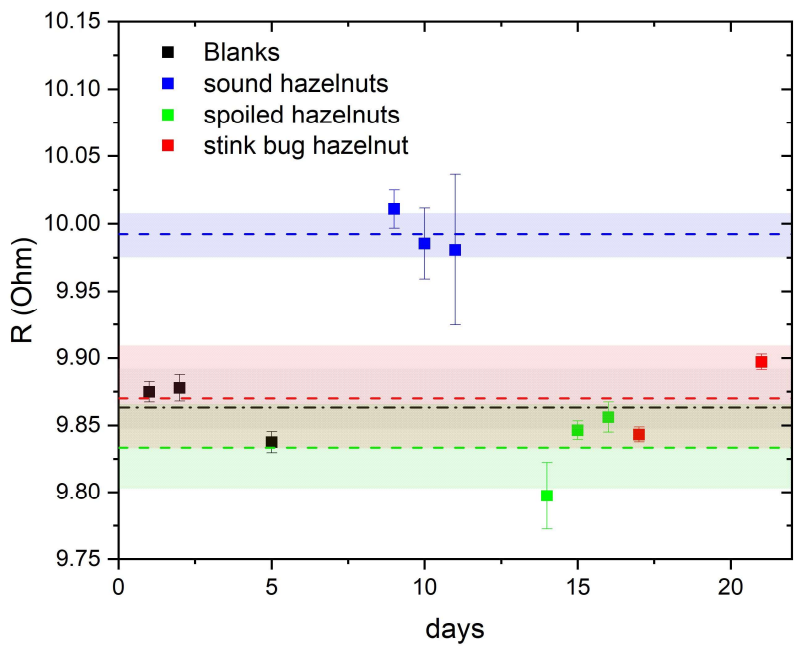

Figure 6: Sensor resistance in a series of measurements spanning more than 20 days. The black points represent blanks, blue points refer to the sensor exposed to sound hazelnuts, green points to stink bug hazelnuts, and red points to spoiled hazelnuts.

Moreover, the device architecture allows a reading of the sensor in real time and a demonstrated long term stability, desirable characteristics for the monitoring of stored hazelnuts during their industrial processing. The sensor gives an averaged evaluation of the stored nuts, and in principle a single sensor can monitor several containers if an appropriate gas sampling is provided. The sensor can be degassed any time the monitored container is changed, to avoid possible interference in the results. In order to quantitatively relate the sensor readout with the hexanal/nonanal concentration, a calibration with a known amount of the target molecules is necessary, even in the perspective of a specific functionalization. Work is in progress to compare the performance of the pristine material with metal-functionalized sensors. We have successfully loaded the porous matrix with gold and palladium nanoparticles, and experiments are ongoing to test and compare sensors with both types of functionalization. A further possibility is to modify the sensor surface with molecular receptors sensitive to the target molecules. The organic functionalization of graphene samples has already been obtained [39, 40], and the development of specific receptors will finally allow to produce sensors with a high degree of specificity.

# 5 Conclusion

We report the successful operation of a sensor able to monitor the conservation state of hazelnuts. The sensor is based on a novel material architecture and realized with a graphene layer epitaxially grown on a porousified crystalline SiC substrate, namely 4H-SiC(0001). The sensor resistance is determined with a lock-in technique in a four–wire configuration setup. The sensor operates at constant temperature, here 40 °C, to avoid the interference of ambient temperature fluctuations with the measurements. In a preliminary investigation, the sensor demonstrated the ability to discriminate between a sample of healthy hazelnuts and samples of harmed hazelnuts, with a high degree of confidence and a signal variation of more than 6 standard deviations between healthy and harmed nuts. This approach offers a good perspective to achieve a commercial device for the monitoring of the rancidity of food and to improve its preservation conditions. Work is in progress to test functionalized sensors and maximize their performance.

**Acknowledgements**

The authors want to acknowledge Soremartec Italia (Ferrero Group) for providing the hazelnut samples and Dr. Valentina Zannier from CNR-Nano for the SEM measurement of the porous material.

# References

[1] United Nation Sustainable Development, https://sdgs.un.org/goals/goal12, **2023**.

[2] W. Schwab, R. Davidovich-Rikanati, E. Lewinsohn, *The Plant Journal* **2008**, *54* 712.

[3] J. Wang, X. Mu, M. Sun, T. Mu, *Applied Materials Today* **2019**, *16* 1.

[4] K. Shrivas, A. Ghosale, P. K. Bajpai, T. Kant, K. Dewangan, R. Shankar, *Microchemical Journal* **2020**, *156* 104944.

[5] E. Singh, M. Meyyappan, H. S. Nalwa, *ACS Applied Materials & Interfaces* **2017**, *9* 34544.

[6] F. Schedin, A. K. Geim, S. V. Morozov, E. W. Hill, P. Blake, M. I. Katsnelson, K. S. Novoselov, *Nature Materials* **2007**, *6* 652.

[7] A. Nag, A. Mitra, S. C. Mukhopadhyay, *Sensors and Actuators A: Physical* **2018**, *270* 177.

[8] N. Chauhan, T. Maekawa, D. N. S. Kumar, *Journal of Materials Research* **2017**, *32* 2860–2882.

[9] C.-F. Wang, X.-Y. Sun, M. Su, Y.-P. Wang, Y.-K. Lv, *Analyst* **2020**, *145* 1550.

[10] L. Ottaviano, D. Mastrippolito, *Applied Physics Letters* **2023**, *123* 050502.

[11] A. Merko¸ci, *2D Materials* **2020**, *7* 040401.

[12] Y. Yan, W. I. Shin, H. Chen, S.-M. Lee, S. Manickam, S. Hanson, H. Zhao, E. Lester, T. Wu, C. H. Pang, *Carbon Letters* **2020**, *31* 177.

[13] S. Venkateshalu, A. N. Grace, *Journal of The Electrochemical Society* **2020**, *167* 050509.

[14] Y. Ping, Y. Gong, Q. Fu, C. Pan, *Progress in Natural Science: Materials International* **2017**, *27* 177.

[15] M. Safarpour, A. Khataee, In S. Thomas, D. Pasquini, S.-Y. Leu, D. A. Gopakumar, editors, *Nanoscale Materials in Water Purification*, Micro and Nano Technologies, 383–430. Elsevier, **2019**.

[16] S. S. Ray, R. Gusain, N. Kumar, In S. S. Ray, R. Gusain, N. Kumar, editors, *Carbon NanomaterialBased Adsorbents for Water Purification*, Micro and Nano Technologies, 225–273. Elsevier, **2020**.

[17] C. McCallion, J. Burthem, K. Rees-Unwin, A. Golovanov, A. Pluen, *European Journal of Pharmaceutics and Biopharmaceutics* **2016**, *104* 235.

[18] A. Tiliakos, A. M. I. Trefilov, E. Tanasă, A. Balan, I. Stamatin, *Applied Surface Science* **2020**, *504* 144096.

[19] R. Thimmappa, M. Gautam, M. C. Devendrachari, A. R. Kottaichamy, Z. M. Bhat, A. Umar, M. O. Thotiyl, *ACS Sustainable Chemistry & Engineering* **2019**, *7* 14189.

[20] T. Wang, D. Huang, Z. Yang, S. Xu, G. He, X. Li, N. Hu, G. Yin, D. He, L. Zhang, *Nano-Micro Letters* **2015**, *8* 95.

[21] X. Shen, T. Sun, L. Yang, A. Krasnoslobodtsev, R. Sabirianov, M. Sealy, W.-N. Mei, Z. Wu, L. Tan, *Nature Communications* **2021**, *12* 820.

[22] S. Veronesi, G. Pfusterschmied, F. Fabbri, M. Leitgeb, O. Arif, D. A. Esteban, S. Bals, U. Schmid, S. Heun, *Carbon* **2022**, *189* 210.

[23] M. Leitgeb, C. Zellner, M. Schneider, M. Lukschanderl, U. Schmid, *J. Electrochem. Soc.* **2018**, *165* E325.

[24] M. Leitgeb, C. Zellner, M. Schneider, U. Schmid, *ECS J. Solid State Sci. Technol.* **2016**, *5* P556.

[25] M. Leitgeb, C. Zellner, C. Hufnagl, M. Schneider, S. Schwab, H. Hutter, U. Schmid, *J. Electrochem. Soc.* **2017**, *164* E337.

[26] A. Macili, Y. Vlamidis, G. Pfusterschmied, M. Leitgeb, U. Schmid, S. Heun, S. Veronesi, *Applied Surface Science* **2023**, *615* 156375.

[27] F. Ozdemir, I. Akinci, *Journal of Food Engineering* **2004**, *63* 341.

[28] B. Baldwin, K. Gilchrist, L. Snare, *Acta Horticulturae* **2005**, *686* 47.

[29] P. Grau, R. Bastias, *Acta Horticulturae* **2005**, *686* 57.

[30] W. Garrone, M. Vacchetti, *Acta Horticulturae* **1994**, *351* 641.

[31] M. B. S. Martin, T. Fernandes-Garcia, A. Romero, A. Lopez, *Journal of Food Processing and Preservation* **2001**, *25* 309.

[32] F. Mencarelli, R. Forniti, D. Desantis, A. Bellincontro, *Ingredienti Alimentari* **2008**, *7*, 39 16.

[33] A. Bellincontro, F. Mencarelli, R. Forniti, M. Valentini, *Acta Horticulturae* **2009**, *845* 593.

[34] A. Das, B. Chakraborty, S. Piscanec, S. Pisana, A. K. Sood, A. C. Ferrari, *Phys. Rev. B* **2009**, *79* 155417.

[35] A. Das, S. Pisana, B. Chakraborty, S. Piscanec, S. K. Saha, U. V. Waghmare, K. S. Novoselov, H. R. Krishnamurthy, A. K. Geim, A. C. Ferrari, A. K. Sood, *Nature Nanotech.* **2008**, *3* 210.

[36] C. Neumann, S. Reichardt, P. Venezuela, M. Drögeler, L. Banszerus, M. Schmitz, K. Watanabe, T. Taniguchi, F. Mauri, B. Beschoten, S. V. Rotkin, C. Stampfer, *Nature Communications* **2015**, *6* 8429.

[37] M. Pedrotti, I. Khomenko, G. Genova, G. Castello, N. Spigolon, V. Fogliano, F. Biasioli, *LWT – Food Science and Technology* **2021**, *143* 111089.

[38] A. Valdés García, R. Sánchez Romero, A. Juan Polo, S. Prats Moya, S. E. Maestre Pérez, A. Beltrán Sanahuja, *Foods* **2021**, *10* 1611.

[39] L. Basta, A. Moscardini, F. Fabbri, L. Bellucci, V. Tozzini, S. Rubini, A. Griesi, M. Gemmi, S. Heun, S. Veronesi, *Nanoscale Adv.* **2021**, *3* 5841.

[40] L. Basta, F. Bianco, A. Moscardini, F. Fabbri, L. Bellucci, V. Tozzini, S. Heun, S. Veronesi, *J. Mater. Chem. C* **2023**, *11* 2630.